\documentclass[10pt,english,pra,reprint,tightenlines]{revtex4-1}
\usepackage[T1]{fontenc}
\usepackage[latin9]{inputenc}
\usepackage{color}
\usepackage{babel}
\usepackage{amsmath}
\usepackage{amssymb}
\usepackage{graphicx}
\usepackage{hyperref}
\hypersetup{colorlinks}

\makeatletter

\usepackage{babel}

\makeatother

\begin{document}
\title{Self-trapped atomic matter wave in a ring cavity}
\author{Jieli Qin}
\email{104531@gzhu.edu.cn; qinjieli@126.com}

\address{School of Physics and Materials Science, Guangzhou University, 230
Wai Huan Xi Road, Guangzhou Higher Education Mega Center, Guangzhou
510006, China}
\author{Lu Zhou}
\email{lzhou@phy.ecnu.edu.cn}

\address{Department of Physics, School of Physics and Electronic Science, East
China Normal University, Shanghai 200241, China}
\address{Collaborative Innovation Center of Extreme Optics, Shanxi University,
Taiyuan, Shanxi 030006, China}
\begin{abstract}
We studied a system of atomic Bose-Einstein condensate coupled to
a ring cavity within the mean-field theory.
Due to the interaction between atoms and light field,
the atoms can be self-trapped. This is verified with both variational
and numerical methods. We examined the stability of these self-trapped
states. For a weakly pumped cavity, they spread during the evolution;
while at strong pumping, they can maintain the shape for a long time.
We also studied the moving dynamics of these self-trapped waves, and
found out that it can be strongly affected by the cavity decay rate.
For a small cavity decay rate, the self-trapped waves undergo a damped
oscillation. Increasing the cavity decay rate will lead to a deceleration
of the self-trapped waves. We also compared the main results
with the semiclassical theory in which atoms are treated as classical particles.
\end{abstract}
\maketitle

\section{Introduction }

Self-trapping is a ubiquitous phenomenon in nature,
for example, solitons in various systems \cite{Drazin1989Solitons,Guo2018Solitons,Kivshar2003,Kono2010Nonlinear,Kevrekidis2008Emergent,CarreteroGonzalez2008Nonlinear},
liquid Helium droplets \cite{Barranco2006Helium} are all some kind
of self-trapped states. The highly tunable atomic Bose-Einstein condensate
(BEC) provides an ideal platform for studying such phenomena. In the
BEC system, the attractive inter-atom interaction (s-wave collision)
results in a Kerr type self-trapping nonlinearity \citep{Dalfovo1999Theory},
and can support bright solitons \citep{Strecker2002Bright,Khaykovich2002Formatioin,Strecker2003Bright}.
If quantum fluctuation (Lee-Huang-Yang correction \cite{Lee1957Eigenvalues})
is included, self-trapped droplets can also be formed in the BEC system \cite{Kadau2016Oberving,FerrierBarbut2016Observation,Chomaz2016Quantum,Cabrera2018Quantum,Cheiney2018Bright}.

Interacting with electromagnetic fields can also lead to nonlinearity
in the BEC systems. When a BEC is illuminated by electromagnetic waves,
it feels a potential from the electromagnetic field. At the same time,
the BEC also serves as a medium, and will backwardly affect the propagation
of the electromagnetic waves. The affected electromagnetic field will
in turn further affect the dynamics of BEC. Due to this feedback mechanism,
nonlinear features arise in the system. Many interesting phenomena
resulting from this type of nonlinearity have been reported \citep{Ritsch2013Cold}.
For atomic gas in a cavity, because of this feedback effect, a dynamical rather than a static
optical lattice is produced. In such a dynamical lattice the atoms feel a friction force,
thus can be cavity cooled and self-organized
\citep{Domokos2001Semmiclassical,Niedenzu2011Kinetic,Ostermann2015Atomic}.
It also soften an optical lattice, and leads to asymmetric
matter wave diffraction \citep{Li2008Matter-Wave,Zhu2011Strong} and
polaritonic soliton \citep{Dong2013Polaritonic}. It also give
rise to phenomena such as spin-exchange \citep{Norcia2018Cavity,Davis2019Photon}
and long-range interactions \citep{Zhang2018Long,Guan2019Two}, self-structuring
\citep{Robb2015Quantum,Ostermann2016Spontaneous},
photon bubble \citep{Mendonca2012Photon,Rodrigues2016Photon}, bistability
\citep{Zhou2009Cavity,Zhou2010Spin,Zhou2011Cavity,Dalafi2017Instrinsic},
spin texture \citep{Landini2018Formation,Ostermann2019Cavity}, chaotic
dynamics \citep{Diver2014Nonlinear}, parametric resonance \citep{Li2019Nonlinear}
in the light-BEC interacting systems. And in the microwave-BEC interacting
systems, soliton \citep{Qin2015Hybrid,Qin2019Tial-free} and vortex
\citep{Qin2016Stable} phenomena have also been reported. Most recently,
it is found that due to such nonlinearity, supersolid can exist in
a driven-dissipative ring-cavity-BEC system \cite{Mivehvar2018Driven,Schuster2020Supersolid},
futhermore a precise gravimeter has been proposed based on the system \citep{Gietka2019Supersolid}.
And a type of novel crystalline droplet has also been predicted in an atom-cavity setup.
\citep{Karpov2019Crystalline}.

Motivated by these progresses, in this work we propose that self-trapped
matter wave can also be supported by the cavity-mediated nonlinearity
in a driven-dissipative cavity-BEC system, and study its stability
and dynamics
using the mean-field theory. In the considered system (see figure \ref{fig:Diagram}),
the cavity light field is built up by transversely illuminating the
BEC, then the built-up light field forms an optical lattice potential
for the BEC. We theoretically demonstrated that this induced optical
lattice can support a self-trapped wave packet. For a weak cavity
pumping, the induced optical lattice is shallow, the localized wave
packet can not be well trapped, therefore it spreads during the time
evolution. And for a strong pumping, the induced optical lattice can
be strong enough to support a long-time stable self-trapped wave packet.
The moving dynamics of these self-trapped waves show very different
features under different cavity decay rates. This is due to the adiabaticity
of the induced optical lattice. The induced optical lattice tends
to follow the movement of the self-trapped wave packet, however, can
not completely catch up. Thus, the self-trapped wave
packet feels a dragging force from the induced optical lattice falling
behind it, therefore it decelerates, and finally stops. When the
self-trapped wave packet stops, the optical lattice also catches up,
and the system reaches a steady state. If the cavity decay rate is
small, even the self-trapped wave packet has been decelerated to the
speed of zero, the optical lattice still can not catch up, so the
self-trapped wave packet will then be accelerated in the opposite
direction. The deceleration and acceleration alternately repeat several
times, and overall the self-trapped wave packet displays a damped
oscillation. In the bad cavity limit \cite{Cirac1995Laser,Horak2000Coherent}
(which means that the cavity decay rate is much larger
than the atoms-cavity coupling, the cavity light field quickly decays
to a steady state, and can instantaneously follow the dynamics of
BEC), the self-trapped wave packet feels no dragging force and will
constantly move with the initial given speed.
At last, we point out that in different atom-cavity setups the phenomena of
self-organization and friction force on atoms
are also predicted by the semiclassical theory in which
the atoms are treated as classical particles
\citep{Niedenzu2011Kinetic,Ostermann2015Atomic}. But here by using the
mean-field theory, the atoms are described by a Schr\"{o}dinger-like
equation, thus the effects of quantum pressure and tunneling of atoms
to neighboring lattice sites can also be included.

The paper is organized as follows: In section \ref{sec:Model}, the
physical model studied in this paper is presented. In section \ref{sec:SolitaryWave},
we show the existence of self-trapped wave packets in the system with
both the variational method and numerical simulation. Examples of
the self-trapped wave packets and their stability are also shown in
this section. In section \ref{sec:MovingDynamic}, the moving dynamics
of the self-trapped wave packets are studied in detail.
And, we briefly compare the main results obtained using mean-field theory
with their semiclassical correspondences in section \ref{sec:Semiclassical}.
At last, the paper is summarized in section \ref{sec:Summary}.

\section{Model\label{sec:Model}}

\begin{figure}
\begin{centering}
\includegraphics{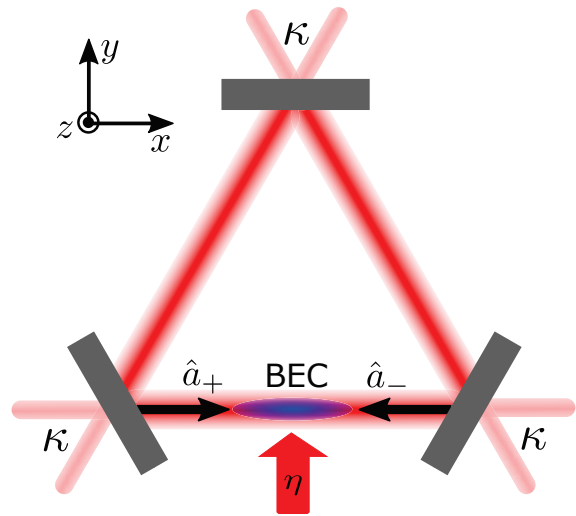}
\par\end{centering}
\caption{Diagram of the considered system. A quasi-one-dimensional atomic BEC
is loaded into a ring-cavity with loss rate $\kappa$. The BEC atoms
interact with two degenerate counter-propagating modes ($\hat{a}_{+}$
and $\hat{a}_{-}$) of the ring-cavity. The system is pumped by transversely
shining a laser on the BEC, the pumping strength is $\eta$. \label{fig:Diagram}}
\end{figure}

We consider a ring cavity-BEC coupling system \citep{Mivehvar2018Driven}
which is schematically shown in figure \ref{fig:Diagram}. A two-level
atomic BEC is trapped along the cavity axis by a tight transverse
confining potential, thus can be reduced to one-dimensional. The atoms
are driven in the transverse direction by an off-resonant (with detuning
$\Delta_{a}$) pump laser, which induces a Rabi oscillation of frequency
$\Omega_{0}$ between the two internal atomic states. The transition
between the two atomic energy levels is also off-resonantly coupled
to the two counter-propagating cavity modes $\hat{a}_{\pm}e^{ik_{c}x}$
($k_{c}$ is the wave number of the cavity modes) with strength $\mathcal{G}_{0}$.
In the far-off-resonant regime $\left|\Delta_{a}\right|\gg\Omega_{0},\mathcal{G}_{0}$,
the excited atomic state can be adiabatically eliminated, and in the
rotating frame of the pump laser, the system can be described by the
following effective Hamiltonian
\begin{equation}
\mathcal{H}=-\hbar\Delta_{c}\left(\hat{a}_{+}^{\dagger}\hat{a}_{+}+\hat{a}_{-}^{\dagger}\hat{a}_{-}\right)+\int\hat{\psi}^{\dagger}H_{a}\hat{\psi}dx,\label{eq:Hamiltonian_eff}
\end{equation}
where the first term describes the two counter-propagating cavity
modes, and the second term accounts for the BEC and its interaction
with the light field. In this equation, $\hbar$ is the Planck constant,
$\Delta_{c}$ is the detuning between the cavity modes and pump laser,
$\hat{\psi}$ is the field operator of the BEC and $H_{a}$ is the
corresponding single-particle Hamiltonian
\begin{align}
H_{a} & =\frac{\hat{p}^{2}}{2m}+V_{ac}+V_{ap},\label{eq:Hamiltonian_atom}
\end{align}
with
\begin{equation}
V_{ac}=\hbar U_0\!\left[\hat{a}_{+}^{\dagger}\hat{a}_{+}+\hat{a}_{-}^{\dagger}\hat{a}_{-}+\left(\hat{a}_{+}^{\dagger}\hat{a}_{-}e^{-2ik_{c}x}+\mathrm{h.c.}\right)\right],\label{eq:V_ac}
\end{equation}
\begin{equation}
V_{ap}=\hbar\eta_0\left(\hat{a}_{+}e^{ik_{c}x}+\hat{a}_{-}e^{-ik_{c}x}+\mathrm{h.c.}\right).\label{eq:V_ap}
\end{equation}
Here, $\hat{p}^{2}/2m$ is the kinetic energy of the BEC atom, $V_{ac}$
is the optical potentials due to two-photon scattering between the
two cavity modes, and $V_{ap}$ is the optical potential due to two-photon
scattering between the pump and cavity modes. The meanings of the symbols
are as follows: $m$ is the mass of the BEC atom, $\hat{p}=-i\hbar\frac{\partial}{\partial x}$
is the momentum operator, $U_{0}=\hbar\mathcal{G}_{0}^{2}/\Delta_{a}$
describes the strength of optical potential $V_{ac}$, and $\eta_{0}=\hbar\mathcal{G}_{0}\Omega_{0}/\Delta_{a}$
is the effective cavity pump strength. In the following contents,
natural unit $m=\hbar=k_{c}=1$ will be applied for simplicity, i.e.,
the length, time, velocity, and energy will be measured in units of
$1/k_{c}$, $m/\left(\hbar k_{c}^{2}\right)$, $\hbar k_{c}/m$, and
$\hbar^{2}k_{c}^{2}/m$.

The BEC usually contains a large number of atoms. To support the
self-trapped wave which is the main subject of this paper, a strong
light field is also needed. Thus, the mean-field approximation
\citep{Zhang2008MeanField} can
be adopted (in reference \citep{Schuster2020Supersolid} which considers
a very similar system, the mean-field results fit the experimental
preservation well). The quantum mechanical operators
can be approximated by their corresponding
mean value c-numbers, $\hat{a}_{\pm}\rightarrow\alpha_{\pm}$ and
$\hat{\psi}\rightarrow\psi$. We further scale $\alpha_{\pm}$
and $\psi$ with the total atom number $N$, i.e., $\alpha_{\pm}\rightarrow\alpha_{\pm}/\sqrt{N}$,
$\psi\rightarrow\psi/\sqrt{N}$. Using such a scaling, the norm of
the wave function $\psi$ becomes
\[
\int\left|\psi\left(x\right)\right|^{2}dx=1.
\]
And we also introduce new parameters $\eta=\sqrt{N}\eta_{0}$ and
$U=U_{0}N$ to account for the many atoms. The equations governing
the dynamics of these mean-field variables can be obtained by taking
the mean values of the corresponding Heisenberg equations

\begin{equation}
i\frac{\partial}{\partial t}\alpha_{\pm}=\left(-\Delta_{c}+U-i\kappa\right)\alpha_{\pm}+UN_{\pm2}\alpha_{\mp}+\eta N_{\pm1},\label{eq:Meanfield_cavity}
\end{equation}
\begin{equation}
i\frac{\partial}{\partial t}\psi=\left[-\frac{1}{2}\frac{\partial^{2}}{\partial x^{2}}+\mathcal{V}_{\mathrm{eff}}\left(x\right)\right]\psi,\label{eq:Meanfield_atom}
\end{equation}
where the cavity loss with rate $\kappa$ has been introduced phenomenologically,
and for conciseness of the equations, here we also defined the following
quantities
\[
N_{\pm1}=\int\left|\psi\left(x\right)\right|^{2}e^{\mp ix}dx,
\]
\[
N_{\pm2}=\int\left|\psi\left(x\right)\right|^{2}e^{\mp2ix}dx,
\]
\[
\mathcal{V}_{\mathrm{eff}}\left(x\right)=\mathcal{V}_{ac}\left(x\right)+\mathcal{V}_{ap}\left(x\right),
\]
\[
\mathcal{V}_{ac}=U\left(\left|\alpha_{+}\right|^{2}+\left|\alpha_{-}\right|^{2}\right)+U\left(\alpha_{+}^{*}\alpha_{-}e^{-2ix}+\mathrm{c.c.}\right),
\]
\[
\mathcal{V}_{ap}=\eta\left(\alpha_{+}e^{ix}+\alpha_{-}e^{-ix}+\mathrm{c.c.}\right).
\]

Letting $\frac{\partial}{\partial t}\alpha_{\pm}=0$ and $\psi\left(x,t\right)=\psi\left(x\right)e^{-i\mu t}$
with $\mu$ being the BEC chemical potential, the steady state of
the system follows equations
\begin{equation}
\mu\psi\left(x\right)=\left[-\frac{1}{2}\frac{\partial^{2}}{\partial x^{2}}+\mathcal{V}_{\mathrm{eff}}\left(x\right)\right]\psi\left(x\right),\label{eq:steadypsi}
\end{equation}
\begin{equation}
\alpha_{+}=-\frac{\left(-\Delta_{c}+U-i\kappa\right)\eta N_{+1}-\eta UN_{+2}N_{-1}}{\left(-\Delta_{c}+U-i\kappa\right)^{2}-U^{2}N_{-2}N_{+2}},\label{eq:alpha_p}
\end{equation}

\begin{align}
\alpha_{-} & =-\frac{\left(-\Delta_{c}+U-i\kappa\right)\eta N_{-1}-\eta UN_{-2}N_{+1}}{\left(-\Delta_{c}+U-i\kappa\right)^{2}-U^{2}N_{-2}N_{+2}}.\label{eq:alpha_m}
\end{align}
Here we point out that equations (\ref{eq:alpha_p}) and (\ref{eq:alpha_m})
also describe the cavity field amplitudes of the system in the bad
cavity limit \cite{Cirac1995Laser,Horak2000Coherent}. In the bad
cavity limit, the cavity light field quickly decays to the steady
state, then $\partial_{t}\alpha_{\pm}\approx0$ can be approximately
applied, so equations (\ref{eq:alpha_p}) and (\ref{eq:alpha_m})
holds. Together with equation (\ref{eq:Meanfield_atom}), dynamics
of the cavity-BEC system in the bad cavity limit can be described.

At last, we see that the dynamics of BEC macroscopic wavefunction
are governed by a Schr{\"{o}}dinger-like equation (\ref{eq:Meanfield_atom}).
This equation is a nonlinear one, since the optical potentials $\mathcal{V}_{ac}\left(x\right)$
and $\mathcal{V}_{ap}\left(x\right)$ felt by the BEC recursively
depend on the wave function $\psi$ of the condensate. This nonlinearity
can support self-trapped waves in the system, which will be discussed
in the next section.

\section{Self-Trapped Matter Wave\label{sec:SolitaryWave}}

In the system, a super-radiation phase transition take place at the
critical pumping strength \citep{Mivehvar2018Driven}
\begin{equation}
\eta_{c}=\sqrt{\frac{\left(-\Delta_{c}+U\right)^{2}+\kappa^{2}}{8\left(-\Delta_{c}+U\right)}.}\label{eq:eta_c}
\end{equation}
Below the critical pumping strength ($\eta<\eta_{c}$), the cavity
light field is almost zero ($\alpha_{\pm}\approx0$), the atoms feel
a negligible optical potential, and will have a uniform distribution.
However, above the critical pumping strength ($\eta>\eta_{c}$), a
considerable intensity of cavity field can be built up, hence the
optical lattice potential acting on the atoms will play a crucial
role. And it will be natural to think that this induced optical lattice
potential can support a self-trapped matter wave packet. This will
be verified by both the variational method analysis and numerical
simulation in the following contents of this section.

\begin{figure}
\begin{centering}
\includegraphics{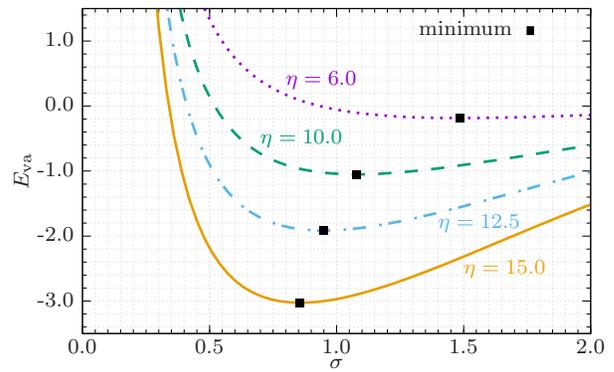}
\par\end{centering}
\caption{Variational Energy. The variational energy $E_{\mathrm{va}}$ is plotted
as a function of the variational parameter $\sigma$ (width of the
wavepacket) in the trial wave function (\ref{eq:psi_va}). The four
lines correspond to different pumping strength $\eta=6.0$, $10.0$,
$12.5$ and $15.0$. The black squares are the minimum points of the
lines. Other parameters used are $U=-0.5$, $\Delta_{c}=-1$, $\kappa=10$.
\label{fig:Variational}}
\end{figure}

To simplify the variational calculation, we further
neglect the terms related to $U,UN_{\pm2}$ (i.e., terms due to two-photon
scattering between the two cavity modes) in equations (\ref{eq:alpha_p})
and (\ref{eq:alpha_m}) under the assumption $\kappa\gg U,UN_{\pm2}$.
And the simplified optical field reads
\begin{equation}
\alpha_{\pm}\approx\frac{\eta N_{\pm1}}{\Delta_{c}+i\kappa}.\label{eq:BadCavity_alpha}
\end{equation}
We see the optical field is determined by the two-photon scattering
between the pump and cavity modes. Then, the ratio between amplitudes
of the two optical potentials $\mathcal{V}_{ac}$ and $\mathcal{V}_{ap}$
is calculated to be
\[
\frac{\mathcal{V}_{ac}}{\mathcal{V}_{ap}}\sim\frac{U\left|\alpha_{\pm}\right|^{2}}{\eta\left|\alpha_{\pm}\right|}=\frac{U\left|N_{\pm1}\right|}{\left|\Delta_{c}+i\kappa\right|}\ll1.
\]
So, compared to $\mathcal{V}_{ap}$, $\mathcal{V}_{ac}$ can be neglected.
Equation (\ref{eq:Meanfield_atom}) which governs the evolution of
atomic BEC can be simplified to the following nonlinear Schr{\"{o}}dinger
equation
\begin{equation}
i\frac{\partial}{\partial t}\psi=-\frac{1}{2}\frac{\partial^{2}}{\partial x^{2}}\psi+\left[\frac{\eta^{2}\left(N_{+1}e^{ix}+N_{-1}e^{-ix}\right)}{\Delta_{c}+i\kappa}+\mathrm{c.c.}\right]\psi.\label{eq:BadCavity_psi}
\end{equation}
The effective Hamiltonian corresponding to this equation can be written
as
\begin{align}
H_{\mathrm{eff}}= & \int\hat{\psi}^{\dagger}\left(x\right)\left(-\frac{1}{2}\frac{\partial^{2}}{\partial x^{2}}\right)\hat{\psi}\left(x\right)dx\nonumber \\
+ & \left[\frac{\eta^{2}}{\Delta_{c}+i\kappa}\left(\int\hat{\psi}^{\dagger}\left(x_{1}\right)e^{ix_{1}}\hat{\psi}\left(x_{1}\right)dx_{1}\right)\right.\nonumber \\
 & \left.\cdot\left(\int\hat{\psi}^{\dagger}\left(x_{2}\right)e^{-ix_{2}}\hat{\psi}\left(x_{2}\right)dx_{2}\right)+\mathrm{h.c.}\right].\label{eq:Heff}
\end{align}
Taking a Gaussian wave packet localized at position $x=0$
\begin{equation}
\psi_{\mathrm{va}}\left(x\right)=\left(\frac{2}{\pi\sigma^{2}}\right)^{1/4}e^{-\left(\frac{x}{\sigma}\right)^{2}},\label{eq:psi_va}
\end{equation}
as the variational trial wave function where the wave packet width
$\sigma$ is the only variational parameter, the variational energy
is integrated to be
\begin{align}
E_{\mathrm{va}} & =\frac{1}{2\sigma^{2}}+\frac{2\Delta_{c}\eta^{2}}{\Delta_{c}^{2}+\kappa^{2}}\exp\left[-\frac{\sigma^{2}}{4}\right].\label{eq:E_va}
\end{align}

In figure \ref{fig:Variational}, the variational energy $E_{\mathrm{va}}$
is plotted as a function of variational parameter $\sigma$ for different
pumping strength $\eta$. We clearly see that there exists a minimal
point on the $E_{\mathrm{va}}\text{{-}}\sigma$ curve, which indicates
the existence of a self-trapped wave packet. And from the figure,
one also expects that the self-trapped wave packet will have a narrower
width under a stronger pumping strength (a larger value of $\eta$),
as the stronger pumping can produce a deeper optical lattice. These
conclusions will be further verified by the numerical simulations.

\begin{figure}
\begin{centering}
\includegraphics{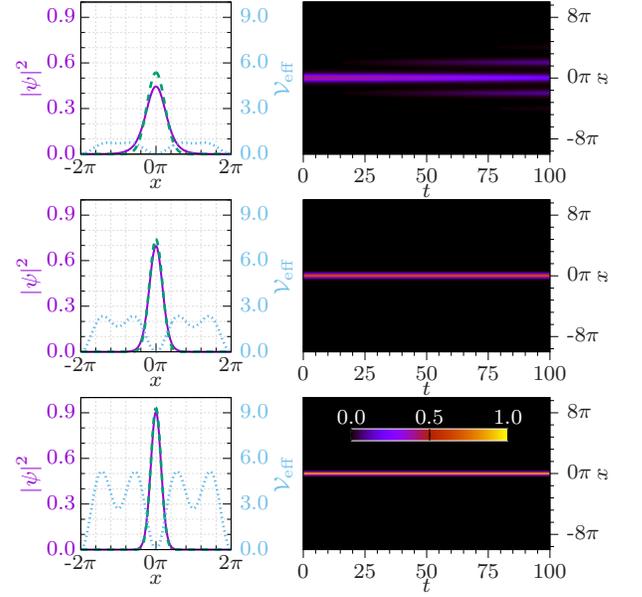}
\par\end{centering}
\caption{Some examples of the self-trapped wave packets and their stability.
Left panels: Density profiles $\left|\psi\right|^{2}$ of the self-trapped
waves and the corresponding induced optical lattice potential $\mathcal{V}_{\mathrm{eff}}$
for pumping strength $\eta=6$ (top panel), $10$ (middle panel),
and $15$ (bottom panel). The violet solid and green dashed lines
are numerical and variational results for $\left|\psi\right|^{2}$
respectively. The cyan dotted lines are numerical results for $\mathcal{V}_{\mathrm{eff}}$.
Right panels: Corresponding time evolution of the self-trapped waves
shown in the left panels. Other parameters are $U=-0.5$, $\Delta_{c}=-1$
and $\kappa=10$. \label{fig:SW}}
\end{figure}

Numerically, the steady state of the system is found by propagating
equations (\ref{eq:Meanfield_atom}, \ref{eq:alpha_p}, \ref{eq:alpha_m})
with the imaginary time method from an initial trial narrow Gaussian
wave packet. Some examples of the numerically found self-trapped wave
packets and their variational counterparts for different pumping strength
$\eta=6$ (top), $10$ (middle), and $15$ (bottom) are shown in the
left panels of figure \ref{fig:SW}, where the induced optical lattice
potentials $\mathcal{V}_{\mathrm{eff}}$ are also plotted. Here the
optical lattices in fact have different bottom energies, but for the
convenience of comparison, we shift all of them to the value of zero.
We see that the variational and numerical results fit each other very
well. And as the pumping strength $\eta$ increases, the depth of
the induced optical lattice also increases, as a result, the width
of self-trapped wave packet deceases. This agrees with our variational
discussion in the previous paragraph.

We also examined the stability of these self-trapped waves by directly
simulating equations (\ref{eq:Meanfield_cavity}) and (\ref{eq:Meanfield_atom}).
The results are shown in the right panels of figure \ref{fig:SW}.
We found that for a weak pumping strength ($\eta=6$ in the top panel),
the induced optical lattice potential is not strong enough to retain
the atoms around a single lattice site, they can tunnel to the neighboring
sites, and the self-trapped wave packet spreads. When the pumping
strength is strong ($\eta=10$, $15$ in the middle and bottom panels),
the self-trapped wave packet can maintain its shape for quite a long
time.

\section{Dynamics\label{sec:MovingDynamic}}

\begin{figure}
\begin{centering}
\includegraphics{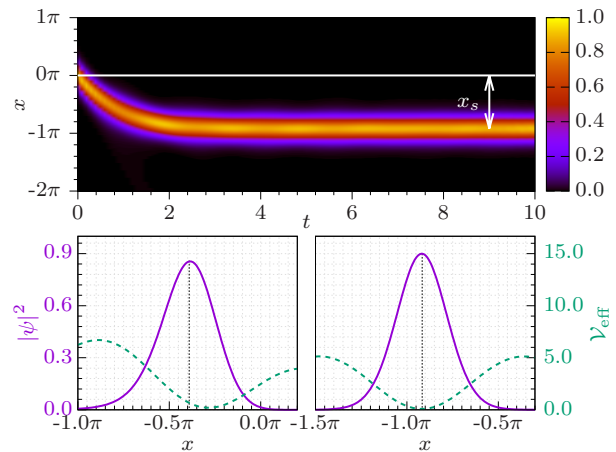}
\par\end{centering}
\caption{Decelerating motion of the self-trapped wave packet. Initially, the
self-trapped wave packet locates at $x=0$ (white solid line), and
its velocity is set to $v_{0}=-3.$ Top panel: The afterward evolution
of the density profile. The limit traveling distance of the wave packet
is $x_{s}=2.88$ (white two-heads arrow). Bottom panels: The density
profiles (violet solid line) and corresponding induced optical lattice
potentials (green dashed line) at $t=0.5$ (left panel) and $t=5.0$
(right panel). The black dotted line is plotted to mark the center
of the wave packet. Other parameters used are $U=-0.5$, $\Delta_{c}=-1$,
$\kappa=10$ and $\eta=15$. \label{fig:Moving}}
\end{figure}

Because of the dissipative nature of the system, the moving dynamics
of the self-trapped waves also show additional features. In the top panel
of figure \ref{fig:Moving}, we initially give the self-trapped wave
packet a velocity of $v_{0}=-3$ by imprinting a phase factor $\exp\left(-iv_{0}x\right)$
on the steady state wave packet \citep{Denschlag2000Generating},
and plot its density profile in the afterward evolution. Unlike the
constant-speed moving of the conventional atomic bright soliton supported
by inter-atom interaction (s-wave collision) \citep{Kevrekidis2008Emergent},
here we see that the self-trapped wave undergoes a decelerating motion.
This is because according to equations (\ref{eq:Meanfield_cavity})
and (\ref{eq:Meanfield_atom}), there is a scope of
timing delay between the change of light field and the moving of atomic
condensate, and the condensate will feel a dragging (friction) force
from the falling behind optical potential \citep{Gietka2019Supersolid}.
As shown in the bottom left panel, at $t=0.5$ the center of the wave
packet has traveled to $x=-1.23$ (black dashed line), but the bottom
of the optical lattice is still left behind at $x=-0.89$, and the
lattice will impede the moving of the condensate. After traveling
some distance, the self-trapped wave packet gets to stop, and the
light field also catches up, thus the system comes back to a steady
sate, see bottom right panel of the figure where the center of the
wave packet and the bottom of the lattice overlap again at $t=5.0$.

This friction force can be used to cool atomic gas
\citep{Domokos2001Semmiclassical,Niedenzu2011Kinetic}.
Moreover, it may also provide an opportunity to simplify
the engineering of the self-trapped waves. For a conventional BEC
bright soliton, if it is required to transfer from one place to another,
one needs to firstly accelerate it, and then one also needs to slow
down and stop it at the destination \citep{Kevrekidis2005Statics,Baines2018Soliton}.
But for the self-trapped waves considered here, the stopping process
can be omitted, one only needs to kick the self-trapped wave packet
with an appropriate initial velocity, then it will travel to and stop
at the destination automatically.

\begin{figure}
\begin{centering}
\includegraphics{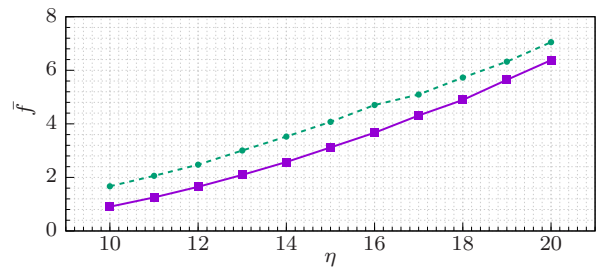}
\par\end{centering}
\caption{Relation between mean friction force $\bar{f}$ and
cavity pumping strength $\eta$.
The squares and circles are data points collected
from mean-field and semiclassical numerical simulations respectively.
The lines are simple linear connections
of the data points to guide the eyes. Parameters used to plot this
line are $U=-0.5$, $\Delta_{c}=-1$, $\kappa=10$ and $v_{0}=-3$.}\label{fig:fVSeta}
\end{figure}

\begin{figure}
\begin{centering}
\includegraphics{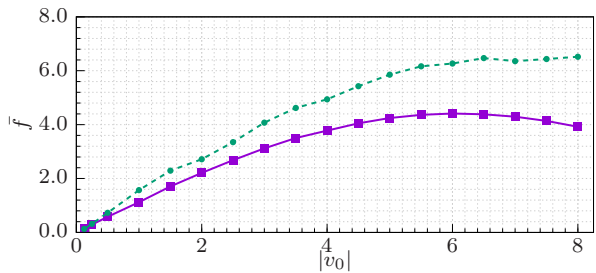}
\par\end{centering}
\caption{Relation between the mean friction force $\bar{f}$ and the initial
wave packet moving speed $v_{0}$.
The squares and circles are data points collected
from mean-field and semiclassical numerical simulations respectively.
The lines are simple linear connections
of the data points to guide the eyes. Parameters used to plot this
line are $U=-0.5$, $\Delta_{c}=-1$, $\kappa=10$ and $\eta=15$.
\label{fig:fVSv0}}
\end{figure}

\begin{figure}
\begin{centering}
\includegraphics{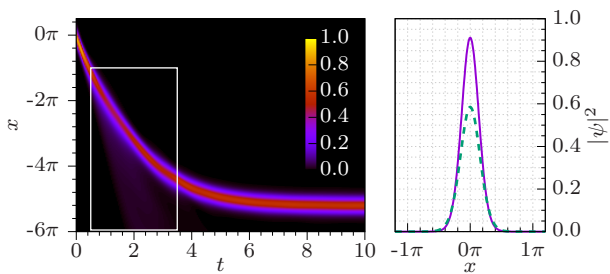}
\par\end{centering}
\caption{Escaping of atoms from a fast-moving self-trapped wave packet. The
initial speed of the wave packet is $v_{0}=-8$. Left panel: Time
evolution of the atomic density $\left|\psi\right|^{2}$. The escaping
of atoms from the self-trapped wave packet is emphasized by the white
box. Right panel: The initial ($t=0$, violet solid line) and final
($t=10$, green dashed line) density profiles of the self-trapped
wave packet. Other Parameters are $U=-0.5$, $\Delta_{c}=-1$, $\kappa=10$
and $\eta=15$. \label{fig:Largev0}}
\end{figure}

Next, we denote the mean friction force felt by the self-trapped wave
as $\bar{f}$, and study its properties in detail. Its value can be
calculated from equation
\begin{equation}
\bar{f}x_{s}=\frac{Nmv_{0}^{2}}{2},\label{eq:meanFriction}
\end{equation}
where we equal the work done by the friction force and the initial
kinetic energy of the self-trapped wave. And here $x_{s}$ is the
limit traveling distance of the wave packet (as shown in figure \ref{fig:Moving}),
which is determined from numerical results.

We firstly examine the dependence of $\bar{f}$ on
pumping strength $\eta$, see figure \ref{fig:fVSeta}. As the cavity
pumping strength $\eta$ increases, the strength of the induced optical
potential also increases accordingly. As can be expected, the self-trapped
wave packet feels a stronger friction force at a larger pumping strength.

In figure \ref{fig:fVSv0}, we plot $\bar{f}$ as a function of the
initial speed $v_{0}$ of the self-trapped wave packet. The faster
the self-trapped wave packet moves, the severer the light field falls
behind, thus the friction force $\bar{f}$ is expected to be proportional
to $v_{0}$. This is numerically observed at small values of $v_{0}$
($v_{0}<3$). However, as $v_{0}$ further increases, the friction
force is saturated; and after $v_{0}>6$ the friction force decrease.
We found that this is caused by the escaping of atoms from the self-trapped
wave packet. When the wave packet moves with a fast speed, a considerable
fraction of atoms can escape from the self-trapped wave packet, thus
the atomic density, and therefore the depth of the optical lattice
is reduced. And a shallower optical lattice will have a weaker friction
effect. This is shown in figure \ref{fig:Largev0}, in the left panel
of which the evolution of atomic density for $v_{0}=8$ is plotted,
and in the right panel the initial and final density profile is compared.
In the figure, the escaping of atoms from the self-trapped wave packet
is characterized by the precursor in the white box. And integrating
the initial and final density profiles, we found that about $27\%$
of the atoms have been lost.

\begin{figure}
\begin{centering}
\includegraphics{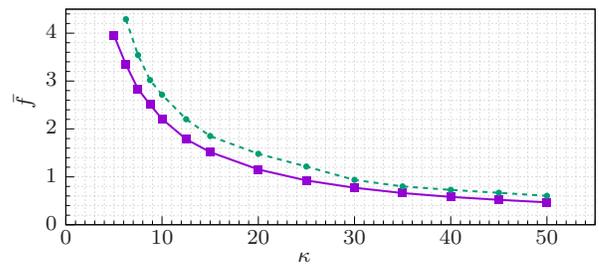}
\par\end{centering}
\caption{Relation between the mean friction force $\bar{f}$ and decay rate
of the cavity $\kappa$.
The squares and circles are data points collected from
mean-field and semiclassical numerical simulations respectively.
The lines are simple linear connection
of the data points to guide the eyes. Parameters used to plot this
line are $v_{0}=-2$, $U=-0.5$, $\Delta_{c}=-1$, and $\eta=15$.\label{fig:fVSkappa}}
\end{figure}

\begin{figure}
\begin{centering}
\includegraphics{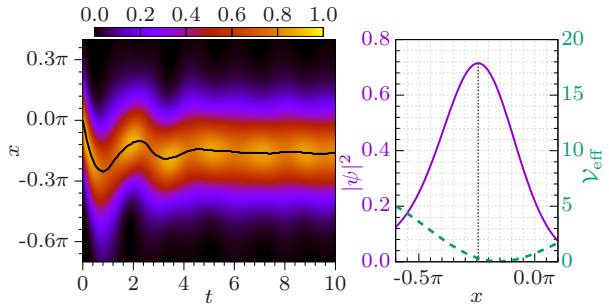}
\par\end{centering}
\caption{Damped oscillation of the self-trapped wave packet under a small cavity
decay rate ($\kappa=2$). Left panel: Time evolution of the atomic
density profile $\left|\psi\right|^{2}$. The black solid line is
the center of the self-trapped wave packet. Right panel: Atomic density
profile (violet solid line) and corresponding optical lattice potential
(green dashed line) at $t=0.96$ (the first time at which the speed
of the wave packet reaches 0, i.e., the first minimal point of the
black line in the left panel). The black dotted line is plotted to
mark the center of the wave packet. Other parameters used are $v_{0}=-2$,
$U=-0.5$, $\Delta_{c}=-1$ and $\eta=3.$\label{fig:DampedOscillatingMotion}}
\end{figure}

\begin{figure}
\begin{centering}
\includegraphics{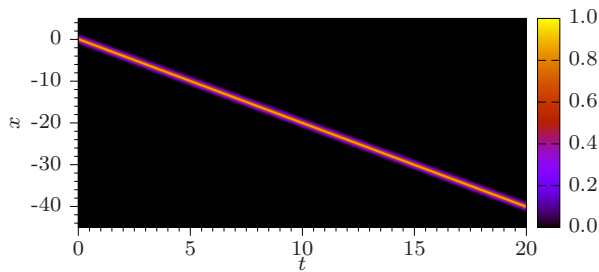}
\par\end{centering}
\caption{Constant-speed moving of the self-trapped wave packet in the bad cavity
limit. To produce this figure, equations (\ref{eq:Meanfield_atom})
together with bad cavity limit optical field formulae (\ref{eq:alpha_p})
and (\ref{eq:alpha_m}) are solved numerically. Parameters used are
$v_{0}=-2$, $U=-0.5$, $\Delta_{c}=-1$, $\eta=75$ and $\kappa=50$.
\label{fig:BadCavityLimit}}
\end{figure}

Since the friction force results from the time delay of the cavity
light field relative to the moving of the atomic matter wave, one
can expect reducing the friction force by shortening the cavity relaxation
time, i.e., increasing the cavity decay rate $\kappa$. This is also
demonstrated by our numerical results, see figure \ref{fig:fVSkappa},
where the mean friction force $\bar{f}$ felt by the self-trapped
wave is plotted as a function of the cavity decay rate $\kappa$.

And we also found that for a small value of the cavity decay rate
$\kappa$, the moving dynamics of the self-trapped waves can show
new features. It undergoes a damped oscillation, as shown in figure
\ref{fig:DampedOscillatingMotion}. In the left panel, the moving
of the density profile (colormap) and center position (black solid
line) of the self-trapped wave is plotted. At the beginning stage,
the delayed light field seriously decelerates the self-trapped wave
packet. Because the light field falls too much behind in this case,
even the speed of the self-trapped wave packet has been decelerated
to zero at $t=0.96$ (the first minimal of the black solid line),
the induced optical lattice potential still can not catch up, see
the right panel where we plot the density profile and the induced
optical lattice potential at this time. As a result, in an afterward
time interval the still falling behind optical lattice accelerates
the self-trapped wave in the opposite direction. Then, the light field
catches up, and decelerates the condensate again. This deceleration-acceleration
process repeats several times, therefore the self-trapped wave undergoes
a damped oscillation.

At last, in the bad cavity limit, the light field can instantaneously
follow the moving self-trapped wave packet, thus it will have no friction
effect on the self-trapped wave. In such a case, we expect that the
self-trapped wave packet will move with its initial speed all the
afterward time. In figure \ref{fig:BadCavityLimit}, the moving of
a self-trapped wave packet is studied under the bad cavity approximation,
i.e., the simulation is done by numerically solving equations (\ref{eq:Meanfield_atom}),
(\ref{eq:alpha_p}), and (\ref{eq:alpha_m}). The expected constant
speed motion is demonstrated by the numerical result.

\section{Comparison to the Semiclassical Theory \label{sec:Semiclassical}}
Taking the atoms as classical polarizable particles, their motion can be approximately
described by a one-dimensional Vlasov equation \citep{Niedenzu2011Kinetic,Ostermann2015Atomic}
\begin{equation}
\frac{\partial f}{\partial t} + v \frac{\partial f}{\partial x}
- \frac{\partial \mathcal{V}_{\mathrm{eff}}}{\partial x} \frac{\partial f}{\partial v} = 0,
\label{eq:Vlasov}
\end{equation}
where $f(x,v,t)$ is the phase space distribution of the atoms with $v$ meaning the velocity.
And the light field is still governed by equation (\ref{eq:Meanfield_cavity}), except
that the variables $N_{\pm 1}$ and $N_{\pm 2}$ are now defined as
\[
N_{\pm 1} = \int \rho(x) e^{\mp i x} dx,
\]
and
\[
N_{\pm 2} = \int \rho(x) e^{\mp 2 i x} dx,
\]
with $\rho(x)$ being the spatial distribution of the atoms
\[
\rho(x,t) = \int f(x,v,t) dv.
\]

\begin{figure}
\includegraphics{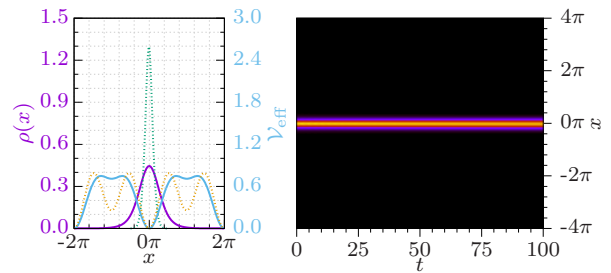}
\caption{
An self-trapped state and its time evolution given by
the semiclassical theory. Left panel: Spatial distribution
(violet and green line)
of the self-trapped state atoms and the corresponding optical
lattice potential (cyan and brown line).
The solid lines are the mean-field result,
while the dotted lines are the semiclassical result. Right
panel: Semiclassical time evolution of the self-trapped state.
The parameters are $U=-0.5$, $\Delta_c=-1$, $\kappa=10$ and
$\eta=6$, which are the the same as in top panels of figure
\ref{fig:SW}. \label{fig:semiClassical}
}
\end{figure}

Such a semiclassical theory also predicts a self-trapped state of the
atoms, see the left panel of figure \ref{fig:semiClassical}.
In this figure, we also plot the
mean-field result for comparison. It is found that the semiclassical
theory gives a narrower spatial distribution of the atoms than the
mean-field theory. This is because the quantum pressure (the kinetic
term in Sch\"{o}dinger equation) which tends to spread
the atomic distribution is absent in the semiclassical theory
[while in the mean-field theory, the atoms are described by the
Schr\"{o}dinger-like equation (\ref{eq:Meanfield_atom}) which
can include the effect of quantum pressure].

Another difference between the mean-field and semiclassical theory results
is the instability of the self-trapped state under weak pumping strength.
Recalling that under weak pumping strength tunneling of atoms to the
neighboring lattice sites will lead to the spreading of self-trapped
state during its evolution (top right panel of figure \ref{fig:SW}).
However, the semiclassical theory treats the atoms as classical particles,
thus the tunneling phenomena can not be included, as a result, it gives
a non-spreading stable evolution of the self-trapped state, see the right
panel of figure \ref{fig:semiClassical} where the parameters
are chosen the same as in the top panels of figure \ref{fig:SW}.

We also compared the mean friction force predicted by
mean-field and semiclassical
theory, as shown in figures \ref{fig:fVSeta}, \ref{fig:fVSv0} and
\ref{fig:fVSkappa}. Because the semiclassical theory
predicts a narrower spatial distribution of the atoms, the
produced optical potential will also be tighter accordingly. And this will
make the semiclassical theory overestimate the friction force.

\section{Summary\label{sec:Summary}}
In summary, we have studied the self-trapped matter waves and their
moving dynamics in a driven-dissipative ring cavity-BEC system
within the mean-field theory. The
self-trapped wave packets have been found by both variational
and numerical methods, and the results fit with each other very well.
The stability of the self-trapped waves is verified by direct numerical
simulations. It is found that for a strong cavity pumping the self-trapped
wave packet can be stable for quite a long time, while for a weak
cavity pumping the self-trapped wave packet suffers a spatial spreading
during its evolution. We also found that the moving dynamics of these
self-trapped waves can be greatly affected by the cavity loss rate.
Three distinct types of motion of the self-trapped waves have been
identified in the system. For a cavity with a small decay rate, the
cavity light field can alternatively drag and push the self-trapped
wave packet, therefore the self-trapped wave packet endures a damped
oscillation. And for a cavity with a moderate decay rate, the self-trapped
wave packet always fells a dragging force from the cavity light field,
and undergoes a decelerating motion. In the bad cavity limit, the
friction force disappears, and the self-trapped wave packet constantly
moves with the initial speed.
The main results are also compared with a semiclassical
calculation where the atoms are treated as classical particles.
We found that the semiclassical theory predicts a narrower
spatial distribution of the atoms, and will overestimate the friction
force. It also misses the instability of the self-trapped state
under weak pumping.
These dynamical tunable self-trapped
waves may find potential applications in fields such as matter wave
interferometers \citep{Polo2013Soliton,McDonald2014Bright,Helm2015Sagnac,Wales2020Splitting}.
\begin{acknowledgments}
This work is supported by the National Natural Science
Foundation of China (Grants No. 11904063, No. 12074120,
No. 11847059, and No. 11374003), and the Science and Technology
Commission of Shanghai Municipality (Grant No.
20ZR1418500).
\end{acknowledgments}

\end{document}